\documentclass{article}

\usepackage[dvips]{epsfig}

\begin{document}

\title{\textbf{Weierstrass $\wp$-lumps}}

\author{RJ Cova\footnote{rcova@luz.ve}  \\ Dept
de F\'{\i}sica FEC \\ La Universidad del Zulia \\ Apartado 15332 \\
Maracaibo 4005-A \\ Venezuela \\
\and WJ Zakrzewski\footnote{W.J.Zakrzewski@durham.ac.uk} \\ 
Dept of Mathematics \\ University of Durham \\
Durham DH1 3LE \\ UK}

\maketitle


\abstract{We present our results of a numerical investigation of the
behaviour of a system of two solitons in the (2+1) dimensional $CP^1$ 
model on a torus. Defined by the elliptic function of Weierstrass,
and working in the Skyrme version of the model, the soliton lumps exhibit 
splitting, scattering at right angles and motion reversal in the various 
configurations considered. The work is restricted to systems with no initial 
velocity.}

\section{Introduction \label{sec:intro}}

Physics in (2+1) dimensions is an area of much active
research, covering topics that include Heisenberg ferromagnets, the quantum
Hall effect, superconductivity, nematic crystals, topological fluids, vortices and
solitary waves \cite{cho}. 
Most of these systems are non-linear. In their
mathematical description the well-known family of sigma models plays a starring
role. One of the simplest models in (2+1) dimensions which is both Lorentz
covariant and which possesses soliton solutions is the $CP^1$ or $O(3)$ sigma
model. Such
solutions are realisations of harmonic maps, by itself a long-established area of 
research in pure mathematics. However, analytical $CP^1$ 
solutions have only been found in the static (2+0) case; their dynamics is
studied using numerical methods and/or other approximation techniques.

Sigma models are also useful as low dimensional analogues of important field
theories in higher dimensions. In fact, the sigma $CP^1$ model in two
dimensional space exhibits conformal invariance, spontaneous symmetry breaking,
asymptotic freedom and topological solitons, properties that resemble some of
those present in a number of forefront field theories in (3+1)  dimensions.
Amongst the latter we have the Skyrme model of nuclear physics \cite{skyrme}.
Initially proposed as a theory of strong interactions between hadrons, it can
now be regarded as a low energy limit of quantum chromodynamics
\cite{witten83}. The Skyrme scheme assumes that its topological solutions
(skyrmions)  correspond, at a classical level, to ground states of light nuclei
with the topological charge (Brouwer degree) representing the baryon number. Of
course, to compare with the properties of real (physical) nuclei one has to add
to the classical results various quantum corrections.

Planar analogues of the Skyrme model involve the addition of some extra terms
to the original $CP^1$ lagrangian in order to stabilise the field solutions
--the `baby skyrmions'.  Without them, the invariance of the pure, planar
$CP^1$ theory under dilation transformations would lead to the instability of
its soliton-lumps. In the traditional approach, where the solitons are harmonic
maps \( \Re_{2} \cup \{\infty\} \approx S_2 \mapsto S_2 \), one adds two terms:  
A Skyrme-like term which controls the shrinking of the lumps and a
potential-like term which controls their expansion. Properly implemented, this
procedure yields stable solitons as confirmed by numerical simulations
\cite{chaos}.

Lately, attention has also being paid to the $CP^1$ model on a torus $T_2$
where the solitons are maps \( T_2 \mapsto S_2 \); this approach amounts to
imposing periodic boundary conditions on the system. A characteristics of this
model is that there are no solitons of topological charge one, a feature
arising because $genus(T_2)$=1.

Recent investigations have unveiled a rich diversity of phenomena in the
toroidal model that goes beyond the two-lump and annular structures one might
expect by analogy with the model on $\Re_{2}$ \cite{non,split,martin}. This
article continues our earlier studies of $CP^1$ skyrmions that were found to
undergo splitting when defined through the elliptic function of Weiestrass
\cite{split}.

\bigskip
In the following section we define the toroidal $CP^1$ model. The numerical
procedure is explained in section \ref{sec:numerical}. Section
\ref{sec:results} reviews previous findings and presents our new results. The
paper closes with section \ref{sec:conclusions}, with some concluding remarks
and suggestions for future research.

\section{The $CP^1$ model on a torus 
\label{sec:themodel}}

Our $CP^1$ skyrmion model is defined by the lagrangian density
%
%
\begin{eqnarray}
{\cal L} &=& \frac{|W_{t}|^{2}-|W_x|^{2}-2 |W_y|^{2}}
               {(1+|W|^{2})^{2}} \nonumber \\
         &-& 2\theta_{1}\frac{(\bar{W_t}W_y-W_{t}\bar{W_y})^2+
                         (\bar{W_t}W_x-W_t\bar{W_x})^2}{(1+|W|^2)^4} \nonumber \\
         &+& 2\theta_{1}\frac{(\bar{W_x}W_y-W_x\bar{W_y})^2}{(1+|W|^2)^4}, 
               \quad x,y \in T_2,
%
%
\label{eq:lagrangian}
\end{eqnarray}
which is the original $CP^1$ lagrangian modified by
the addition of a  Skyrme, $\theta_1$-term \( (\theta_1 \in \Re) \). 
We adopt the notation \( W_t=\partial_t W, W_x=\partial_x W, 
W_y=\partial_y W \) for the derivatives; the bar denotes complex conjugation. 

In order to obtain stable lumps on the torus it is sufficient to supplement the
pure $CP^1$ lagrangian with a $\theta_1$ term as shown in equation
({\ref{eq:lagrangian}). This is also the case in the (3+1) dimensional Skyrme
theory. As remarked in the introduction, in order to stabilise the lumps in the
traditional $CP^1$ model on $\Re_{2}$ we have to add, other than the $\theta_1$
term, also a potential-like term.

Also it is worth mentioning that on $T_2$ one no longer has the problem
confroted in the extended plane, whose non-compactness brings about formal
difficulties in defining the metric on the moduli space of static soliton
solutions \cite{martin}.

\bigskip
For the $CP\sp1$ model to be defined on a torus we require  the complex field $W$ 
to obey the periodic boundary conditions 
%
%
\begin{equation}
W[z+(m+in)L]=W(z), \qquad \forall t,
\label{eq:boundary}
\end{equation}
where $m,n=0,1,2,...$ and  $L$ is the size of the torus. The static
solitons (instantons) are elliptic functions which may be written as
%
%
\begin{equation}
W=\lambda \, \wp(z-a) + b, 
\qquad \lambda, a, b \in {\cal Z},
\label{eq:w}
\end{equation}
where $\wp$ denotes the elliptic function of Weierstrass. The partial 
fraction representation of $\wp$ reads
%
\begin{equation}
\wp(u)=u^{-2} - \sum_{-\infty}^{\infty}\{[u-(m+in)L]^{-2} - [(m+in)L]^{-2}\},
\label{eq:p}
\end{equation}
the summation being over the integers $m,n$ excluding the
combination $m=n=0$. A comprenhensive treatment of elliptic functions can be
found in \cite{goursat,lawden}. 

The function (\ref{eq:p}) is of the second order, hence (\ref{eq:w}) represents
solitons of topological charge 2.  A particularity of our instantons is that
they have no analytic representative of charge one (the model on $\Re_{2}$ has
representatives in all topological classes).  It is important to note that
(\ref{eq:w})  is an approximate solution of the model (\ref{eq:lagrangian}),
and it becomes an exact static solution in the $O(3)$ limit ($\theta_1$=0)
where it satisfies the ensuing field equation. This means that the $CP^1$ lumps
should evolve only for $\theta_1 \neq 0$.

\bigskip
Our first investigations \cite{non} of periodic solitons involved the use of
the Weierstrass pseudo-elliptic function $\sigma$. Then $W$ was taken to be
given by:
%
%
%
\begin{equation}
W=\lambda \prod_{j=1}^{\kappa} \frac{\sigma(z-c_{j})}{\sigma(z-d_{j})},
\quad \sum_{j=1}^{\kappa} c_{j}=\sum_{j=1}^{\kappa} d_{j},
\label{eq:wsigma}
\end{equation}
where the accompanying selection rule between the zeroes ($c_j$) and poles ($d_j$)
guarantees that $W$ is elliptic. 

Observe that when $\kappa=1$ in (\ref{eq:wsigma}) the constraint between the
zeroes and the poles must be relaxed lest a trivial configuration $W$ is
desired. Although this procedure renders $W$ pseudo-periodic, starting with
such a field a periodic ansatz of topological carge one was constructed in
\cite{non}. This gave a field which was an approximate solution of the
equations of motion for $\theta_1 \neq 0$, but became singular as $\theta_1
\rightarrow 0$.

The power series for $\sigma$ on a square torus may be cast into the form
%
\begin{equation}
\sigma(u)=\sum_{j=0}^{\infty}{G}_{j}u^{4j+1},
\qquad G_j \in \Re.
\label{eq:sig}
\end{equation}
With the help of equation (\ref{eq:p}) the coefficients $G_j$ can be 
calculated by expanding 
$$
\sigma(u)=\int_{0}^{u} [\wp(v)-1/v^2].
$$
In general, the coefficients are written in terms of the so-called
\emph{invariants} $g_2(L)$ and $g_3(L)$. However, square boundary conditions
effectively set $g_3=0$, the lemniscate case, leading to a relatively simple
expression of the form (\ref{eq:sig}).

\bigskip
Let us also add that we can re-express the $\wp$ function through $\sigma$ 
via the formula
%
\begin{equation}
\wp(u)=-\frac{d^2}{du^2}\ln[\sigma(u)],
\label{eq:psig}
\end{equation}
and then perform the computation of $\sigma$ with the same numerical subroutine
which was used in reference \cite{non}. 

\bigskip
Note that each factor $\sigma(u)/\sigma(v)$ in the field (\ref{eq:wsigma})  
can be used to represent a single soliton, providing a setting to studying more
or less independent lumps in all topological classes. Solitons in the
$\sigma$-picture may, for example, be boosted independently. On the other hand,
through equation (\ref{eq:w}) we can only have solitons of even topological
index because $\wp$ is itself an elliptic function of order two. Lumps in the
$\wp$-picture are less independent than their siblings in the
$\sigma$-formulation, for changing the values of the parameters in (\ref{eq:w})
always affects \emph{both} lumps. Strictly speaking, however, truly independent
solitons can only be obtained in the asymptotic regime of large lump
separation, which really never happens on a compact manifold \cite{private}.

\bigskip
Now let us display some useful relations satisfied by the $\wp$ function,
relations which follow from (\ref{eq:p}). They are:
%
\begin{equation}
\wp(-u)=\wp(u), \quad \wp(iu)=-\wp(u), \quad \wp(\bar{u})=\overline{\wp(u)}.
\label{eq:properties}
\end{equation}
Moreover, there exists a useful algebraic relation between $\wp$ and $d\wp/du$ 
on a square torus which reads
%
%
\begin{equation}
[\frac{d\wp(u)}{du}]^2=4 \wp(u)[\wp(u)^2-\wp(L/2)].
\label{eq:id}
\end{equation}

It can be deduced from the equations (\ref{eq:properties}) and (\ref{eq:id})
that the function $\wp$ is purely imaginary on the diagonals bisecting the
fundamental cell and real on the central cross and on the boundary of the cell:
%
\begin{equation}
\wp=
\left\{
\begin{array}{ll}
  \mbox{imaginary} : & \mbox{on central diagonals};
  \\
  \\
  \mbox{real} : & \mbox{on central cross and boundary}.
\end{array}
\right.
\label{eq:central}
\end{equation}

The static energy density associated with the field
(\ref{eq:w}) can be read-off from the lagrangian density (\ref{eq:lagrangian}). 
Using formula (\ref{eq:id}) we may write
%
\begin{equation}
E=e(1+4\theta_1 e), \quad
e=8\lambda^2 \frac{|\wp(z-a)||\wp^{2}(z-a)-\wp^{2}(L/2)|}
                        {[1+|\lambda\wp(z-a)+b|^2]^2}.
\label{eq:energydensity}
\end{equation}
Pictures of $E$ reveal a distribution of lumps localised in space.
The parameter $\lambda$ is related to the size of the lumps, $b$
determines their mutual separation and the parameter $a$ merely shifts
the system as a whole on the torus. Throughout our simulations we
have chosen the values
$$
\lambda=(1,0), \quad a=(2.025, 2.05), \quad \theta_1=0.001 \;(\mbox{or zero}),
$$
and studied different configurations by varying $b$.
                                                        
\section{Basic numerical set up \label{sec:numerical}}

We have taken fields of the form (\ref{eq:w}) as the initial conditions for our
time evolution, studied numerically. Since the field $W$ may become arbitrarily
large, we have preferred to run our simulations in the $\phi$-formulation of
the model. Its field equation follows from the lagrangian density
(\ref{eq:lagrangian}) with the help of the stereographic projection
\begin{equation}
\vec{\phi}=(\underbrace{\frac{W+\bar{W}}{|W|^{2}+1}}_{\phi_1},
\underbrace{i\frac{-W+\bar{W}}{|W|^{2}+1}}_{\phi_2},
\underbrace{\frac{|W|^{2}-1}{|W|^{2}+1})}_{\phi_3},
\label{eq:wphi}
\end{equation}      
with the real scalar field $\vec{\phi}$ satisfying \( \vec{\phi}.\vec{\phi}=1 \). 
One has
%
\begin{eqnarray}
\lefteqn{\vec{0}=
(\partial^{\mu}\partial_{\mu}-
\vec{\phi}.\partial^{\mu}\partial_{\mu} \vec{\phi}) \vec{\phi} }
\nonumber \\
 &+&
2 \theta_{1} [ \partial^{\mu}\partial_{\mu} \vec{\phi}
(\partial^{\nu} \vec{\phi} . \partial_{\nu} \vec{\phi})+
\partial_{\nu} \vec{\phi}(\partial_{\mu}\partial^{\nu} \vec{\phi}
. \partial^{\mu} \vec{\phi})-
\partial_{\nu}\partial_{\mu} \vec{\phi} 
(\partial^{\nu} \vec{\phi} . \partial^{\mu} \vec{\phi}) \nonumber \\
 &-&
\partial^{\mu} \vec{\phi} (\partial_{\nu}\partial^{\nu} \vec{\phi}
. \partial^{\mu} \vec{\phi})+
(\partial^{\mu} \vec{\phi} . \partial_{\mu} \vec{\phi})
(\partial^{\nu} \vec{\phi} . \partial_{\nu} \vec{\phi}) \vec{\phi}
\nonumber \\  
 &-&
(\partial_{\mu} \vec{\phi} . \partial_{\nu} \vec{\phi})
(\partial^{\mu} \vec{\phi} . \partial^{\nu} \vec{\phi}) \vec{\phi} ],
\label{eq:skyrmet2inphi}
\end{eqnarray}
where $\mu, \nu = 0,1,2$ are the Lorentz indices; as usual we have 
\( t,x,y=x^0,x^1,x^2 \).

We compute the series (\ref{eq:sig}) up to $G_5$, the coefficients $G_j$ being
in our case negligibly small for $j > 5$:

\begin{displaymath}
\left.
\begin{array}{lll}
G_{0}&=&1 \\
G_{1}&=&-0.7878030 \\
G_{2}&=&-0.221654845 \\
G_{3}&=&9.36193 \times 10^{-3} \\
G_{4}&=&7.20830 \times 10^{-5} \\
G_{5}&=&2.37710 \times 10^{-5}
\end{array}
\right\}.
\end{displaymath}             

 We have employed the fourth-order Runge-Kutta method and approximated the
spatial derivatives by finite differences.  The laplacian has been evaluated
using the standard nine-point formula and, to further check our results, a
13-point recipe has also been used. Respectively, the laplacians are:

\begin{displaymath}
\nabla^2 = \frac{\left[
 \begin{array}{cccc}
  1 & 4   & 1  \\
  4 & -20 & 4 \\
  1 & 4   & 1
\end{array}
\right]}
{6 \times a^2},
\end{displaymath}

\begin{displaymath}
\nabla^2= \frac{\left[
 \begin{array}{ccccc}
    &    & -1  &    &   \\
    & 1 & 12 & 1 &    \\
  -1 & 12 & -48  & 12 & -1  \\
    & 1 & 12 & 1 &   \\
    &    & -1  &    &
\end{array}
\right]}
{10 \times a^2}.
\end{displaymath}          

The discrete model has been evolved on a \( n_x \times n_y = 200 \times 200 \)
periodic lattice with spatial and time steps $\delta x$=$\delta y$=0.02 and
$\delta t$=0.005, respectively. The vertices of the fundamental lattice we have
used for our simulations were at
\begin{equation}
(0,0),\; (0,L),\; (L,L),\; (L,0), \quad L=n_x \times \delta x=4.
\label{eq:cell}
\end{equation}

Unavoidable round-off errors have gradually shifted the fields away from the
constraint \( \vec{\phi}.\vec{\phi}=1 \). So we rescale 
$$
\vec{\phi} \rightarrow \vec{\phi}/\sqrt{\vec{\phi}.\vec{\phi}}
$$
every few iterations. 
Each time, just before the rescaling operation, we evaluate the quantity \(
\mu \equiv \vec{\phi}.\vec{\phi} - 1 \)\, at each lattice point. Treating the
maximum of the absolute value of $\mu$ as a measure of the numerical errors,
we find that max$|\mu|$ $\approx$ 10$^{-8}$.  This magnitude is useful as a
guide to determine how reliable a given numerical result is. Usage of an
unsound numerical procedure in the Runge-Kutta evolution shows itself as a
rapid growth of max$|\mu|$; this also occurs, for instance, in the 
limit $\theta_1 \rightarrow 0$ when the unstable energy lumps 
become infinitely spiky.

\section{Results \label{sec:results}}

In reference \cite{split} we considered two cases: $b$=(0,0) and $b$=(1,0).  
For \fbox{$b$=(0,0)} one has a most evenly spread energy distribution.  
Notably, this homotopy-two class configuration exhibits \underline{four lumps}
(rather than two) sitting on the central diagonals of the basic grid
(\ref{eq:cell}). (See the top-left graph of figure \ref{fig:pfone}.) When
evolved from rest in the Skyrme scheme, this quartet has moved along the
diagonals under the action of a net repulsive force. They have evolved in a
manner that resembled a scattering at right angles as illustrated in upper half
of figure \ref{fig:pftwo}. No splitting has been observed.

The state \fbox{$b$=(1,0)} corresponds to a couple of lumps placed along the
cordinate axes as seen in figure \ref{fig:pfone} (top-right). This is certainly
a more familiar picture for solitons belonging to a charge-two topological
sector. Unfamiliar, though, is their novel dynamics: starting from rest after a
while each soliton splits into two lumps.  The split is in the direction
perpendicular to the line joining the solitons.  As time goes by, the offspring
skyrmions glue back together, split again and so forth. This is depicted in the
lower part of figure \ref{fig:pftwo} and is the splitting phenomenon described
in the introduction. The time at which the lumps begin to split is $t \approx$
7.

The lower half of figure \ref{fig:pfone} shows plots of \(E_{max} \, vs. \, t\)
for the above cases, including the unstable $O(3)$ case. In the latter, as
expected, the lumps do not move at all (let alone split) with the passing of
time --as long as the initial speed is zero. This is in accordance with our
expectations, for, as we recall, the field (\ref{eq:p}) is a static solution of
the pure $O(3)$ model.

\bigskip
Note that in general $b=(\alpha,0)$ corresponds to solitons initially located 
on the central cross of the grid. If $\alpha > 0$ ($\alpha < 0$) the lumps lie on
the vertical (horizontal) axis. Our qualitative
results are unaffected by reasonable values of $\alpha$.      
In connection with these configurations it is interesting to
consider the case where the simulations run in the Skyrme format up
to $t=t_0$, after which they evolve with $\theta_1$=0. 
In other words, we perform the simulations with a Skyrme field as the initial
condition for the $O(3)$ evolution. Since the splitting forces act only when
$\theta_1 \neq 0$, we expect that the larger $t_0$ is the 
sooner the solitons will begin to divide up.  

Let us analise the situation arising from \fbox{$b=(-1,0)$}, which positions
the lumps along the horizontal axis. Figure \ref{fig:pf3} illustrates the
evolution of the system for the sample cases $t_0$=1.3, 1.5. We find that after
splitting the extended structures reunite but eventually break up for good. In
agreement with our estimates, for $t_0$=1.3 the system splits later than it
does for $t_0$=1.5. Also apparent is that the lumps show a stronger tendency to
glue back in for $t_0$=1.3. We may now compare these plots with figure
\ref{fig:pf4}, which exhibits the splitting situation for $b=(-1,0)$ in the
limit $t_0=\infty$, \emph{i.e.,} when the routine evolves with the Skyrme term
on at all times.

The previous diagrams suggest that a $t_0$ might exist for which the lumps
would come back together without further division. After some trial and error
we have found that such critical time is approximately $t_0 \approx$ 1.25. The
plots presented in figure \ref{fig:pf5} show that the forces brought about by
the skyrmionic initial conditions are just enough to set the solitons
oscillating in a break-up-join-up fashion. The skyrmions eventually settle
together and stay that way for as long as the numerical procedure can be
trusted: In the pure format $E_{max}$ shrinks indefinitely as usual although it
takes quite a while for the system to blow up ($t \approx$ 100).
This soliton is almost a static solution of the model.

Let us point out that when defined via equation (\ref{eq:wsigma}) the periodic
chunks of energy were found not move at all when $v_0=0$, neither in the pure
case nor in the Skyrme case \cite{non}. Such result is rather unexpected for
skyrmions, for they are only approximate solutions of the field equation.
Therefore, solitons obeying (\ref{eq:boundary}) look sensitive to the choice of
function describing them. Not so on $\Re_{2}$, where the qualitative behaviour
of the solitons is qualitatively the same for all the choices.
 
\bigskip Let us now study \fbox{$b$=(0,1)}. For this value the energy
distribution (\ref{eq:energydensity}) associated with (\ref{eq:w}) has the form
of two lumps placed on a bisecting diagonal of the elementary grid (figure
\ref{fig:pf6}). Our numerical simulations show that the skyrmions attract each
other and collide at the centre of the grid. They coalesce indistinguishably
for a moment and then re-emerge perpendicularly to the initial line of
approach, scattering off at 90$^{\circ}$. Afterwards they continue towards the
corners (which are the same point in a periodic set-up like ours) where they
again scatter off at 90$^{\circ}$. These events are depicted in figures
\ref{fig:pf7} and \ref{fig:pf8}; the corresponding $E_{max}(t)$ diagram is the
Skyrme curve shown in the lower half of figure \ref{fig:pf9}.  Such
multi-scattering process goes on indefinitely, and no splitting is observed.

We have also evolved the $b=(0,1)$ case with $\theta_1=0$. This time the energy
chunks have not moved at all with the passing of time until they have become too
narrow and the numerical procedure has broken down (see figure \ref{fig:pf9}).
The shrinking of solitons and the problem of singularity formation in this
model was predicted in \cite{martin}, using the geodesic approximation.  

\bigskip
Next we consider \fbox{$b$=(1,1)}. Here we have a pair of lumps initially
situated at some almost random points, neither on the central cross nor on the
central diagonals of the lattice.  Figures \ref{fig:pf10} and \ref{fig:pf11}
reveal an interesting evolution for these skyrmions. As time elapses they move
along the flat torus along the $x$-axis, disappearing/re-appearing through the
lateral edges of the basic cell (\ref{eq:cell})  in a continuos, periodic
motion. For instance, the skyrmion at the upper (lower) half of the grid
journeys towards decreasing (increasing) $x$ whilst keeping its $y$ coordinate
roughly constant. The lump disappears into the line $x=0$ ($x$=4) re-emerging
from the opposite side $x=4$ ($x=0$). By imagining the flat torus as the
product manifold of two circles, \( T_2 = S_1 \times S_1 \), we can visualise
the motion just described as the trajectory on a circle with our $x$ coodinate
corresponding to an angle.

The two lumps continue moving along and, around $t$=69.5, they reverse their
motion: The skyrmion at the upper (lower) half of the grid starts towards
increasing (decreasing) $x$ whilst keeping its $y$ coordinate roughly constant.
Now the upper (lower) lump will disappear through the line $x=4$ ($x$=0) and
re-emerge on the opposite side $x=0$ ($x=4$). Figure \ref{fig:pf12} presents
the variation of the coordinates in time, as well as the evolution of the peak
of the energy density. No splitting was observed for fields of the type
$b$=(1,1).
  
\bigskip
Thus, for skyrmions whose initial positions are determined by 
$b=(\alpha,\beta)$ our results indicate that 
\begin{itemize}
\item \textbf{splitting only occurs} for \( b=(\alpha,0), \, \alpha \neq 0 \), 
that is, when the solitons originally lie on the central cross of the network
(\ref{eq:cell}). A repulsive, splitting force acts \emph{within} each lump. By
using a Skyrme field as an initial condition for the $O(3)$ evolution, it is
possible to introduce some attractive forces so that the lumps stick back
together.
\item \textbf{Head-on right angle scattering only takes place} for 
\( b=(0,\beta) \), that is, when the skyrmions initially sit on the central
diagonals of the grid. There is a net attractive force between the lumps. Now,
 \begin{itemize}
  \item if \( \beta \neq 0 \) then we have two lumps located on a diagonal;
  \item if \( \beta = 0 \) then we have four lumps symmetrically situated on both
          main diagonals. In this case, the scattering becomes a
          breather-like vibration with the four lumps moving in towards the centre
          of the grid and moving out again. Interestingly, a look
          at the property (\ref{eq:central}) indicates that for $b$=(0,0) 
          the soliton $W=\wp(z-a)$ is purely imaginary.  
  \end{itemize}
\item \textbf{Neither splitting nor scattering} is observed when 
\( b=(\alpha,\beta), \, \alpha, \beta \neq 0 \). Here we have a configuration of 
two energy humps propelled along the lines parallel to the edges of the cell and, 
later on, experiencing
a 180$^{\circ}$ change in their sense of motion. This looks like scattering 
at a large impact parameter. We remark that all systems considered in the present
paper start off from rest.
\end{itemize}

%
\bigskip

The right angle scattering is a generic feature of sigma-type models
\cite{rosen}. In the case of the $CP^1$ dynamics on $T_2$ the scattering at
90$^{\circ}$ was theoretically hinted at in \cite{martin} (pure case),
where the initial value problem was defined in terms of the $\wp$ function
(\ref{eq:p}); the solitons were evolved using the geodesic approximation.  In
reference \cite{ejp} we evolved the $\wp$ solitons using a numerical simulation
of the full $CP^1$ model (pure case) and confirmed the 90$^{\circ}$
scattering predictions
of \cite{martin}. Earlier on \cite{non}, we had observed the said scattering on
$T_2$ (for both the pure and Skyrme cases) with solitons expressed through the
$\sigma$ function (\ref{eq:sig}). The present work further establishes these
results, with the added interesting feature that the lumps, in the Skyrme case, 
collide and scatter off even though they have no initial speed.

Among other sigma models that exhibit $90^{\circ}$ scattering we can mention
monopoles \cite{atiyah} and vortices \cite{samols}.
                 
\bigskip
Finally, it is worth to note from the bottom-left diagram of figure
\ref{fig:pfone} that the lumps for the $b=(0,0)$ case shrink very little, which
suggests quite a stable configuration. So we have performed a number of
simulations with $b$=(0,0) in the pure $O(3)$ format. Indeed, as figure
\ref{fig:pf13} shows, the solitons shrink at a very slow rate and can be
regarded as stable for practical purposes. Contrary to the case of $\theta_1=0$
processes with no initial speed, a displacement in this four-lump configuration
is observed, one of breather-like characteristics.

\begin{figure}
\mbox{\epsfig{file=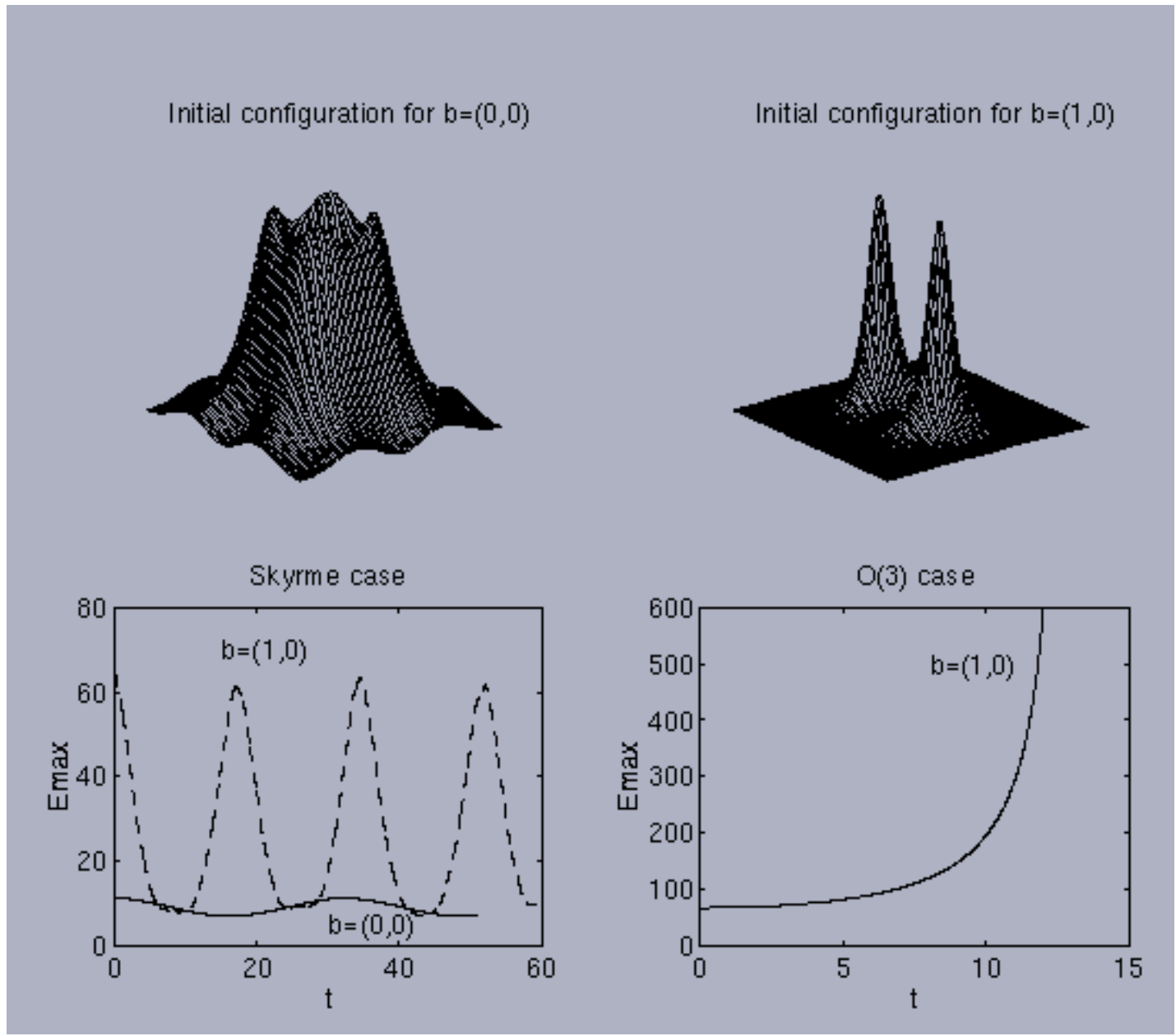}}
\caption{Energy density configurations at $t=0$ and the evolution of the
maximum $E_{max}$. Above we see the special case featuring four lumps
[$b$=(0,0)] and the more common situation with a pair of lumps [$b$=(1,0)]. The
evolution of $E_{max}$ is illustrated bellow, both for the Skyrme and pure
$O(3)$ case. The instability of the latter is manifest by the lumps shrinking
non-stoppingly.}
\label{fig:pfone}
\end{figure}

\begin{figure}
\mbox{\epsfig{file=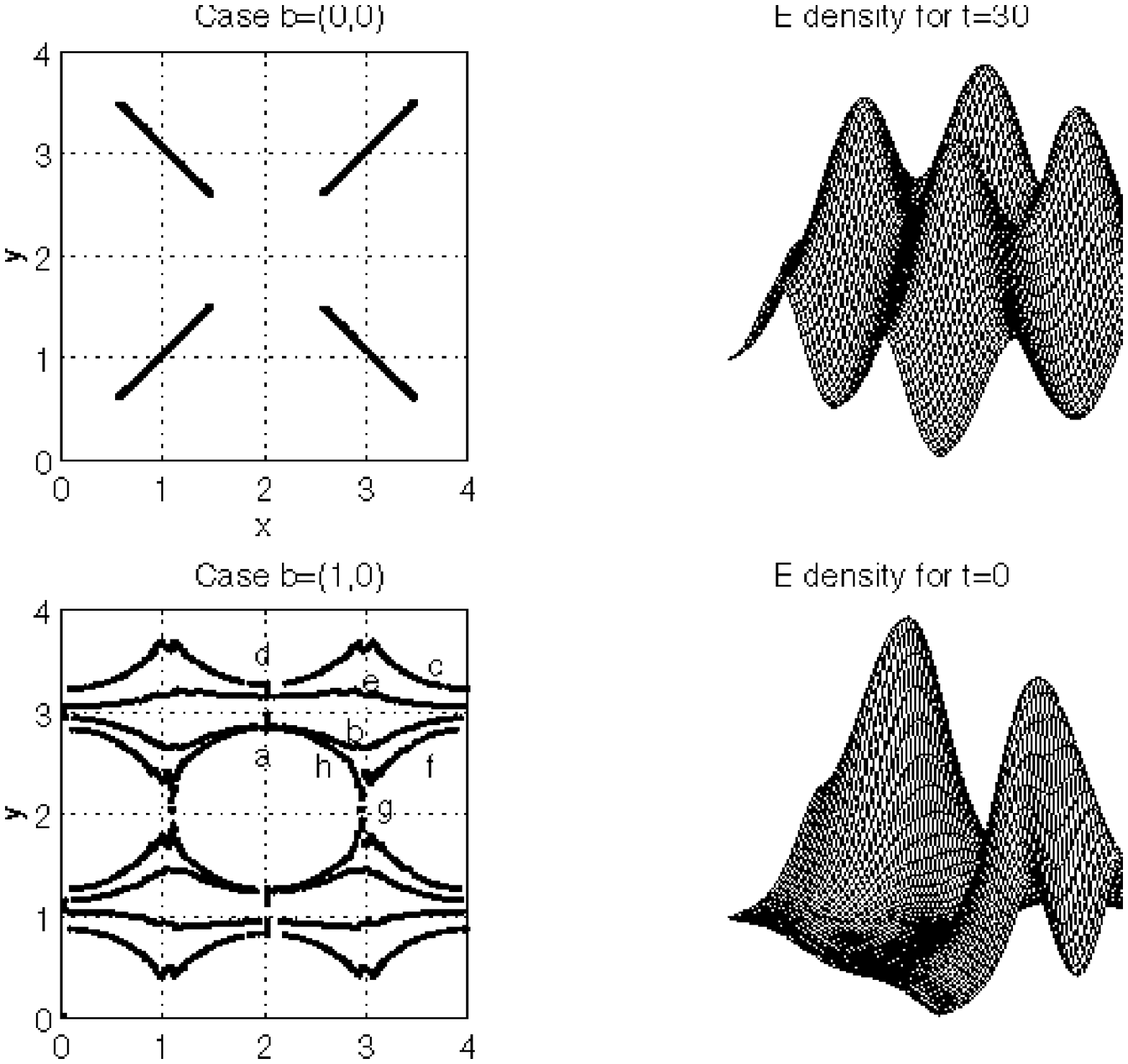}}
\caption{\underline{Above:} Skyrmion for the exceptionally symmetric
case $b$=(0,0). The four lumps stay on the diagonals and move towards 
the corners and coalesce. They break-off again and proceed back to 
the centre of the lattice at $t$=51. \underline{Below:} The
$b$=(1,0) skyrmions split each in two lumps that transit complicated paths;
the labels $a-h$ refer to one of the `half-lumps'. The $t$=30
picture features the situation shortly after the `fractional' lumps
reunite at $d$ (and at its symmetrical point) and begin to travel
centrewards.}
\label{fig:pftwo}
\end{figure}

\begin{figure}
\mbox{\epsfig{file=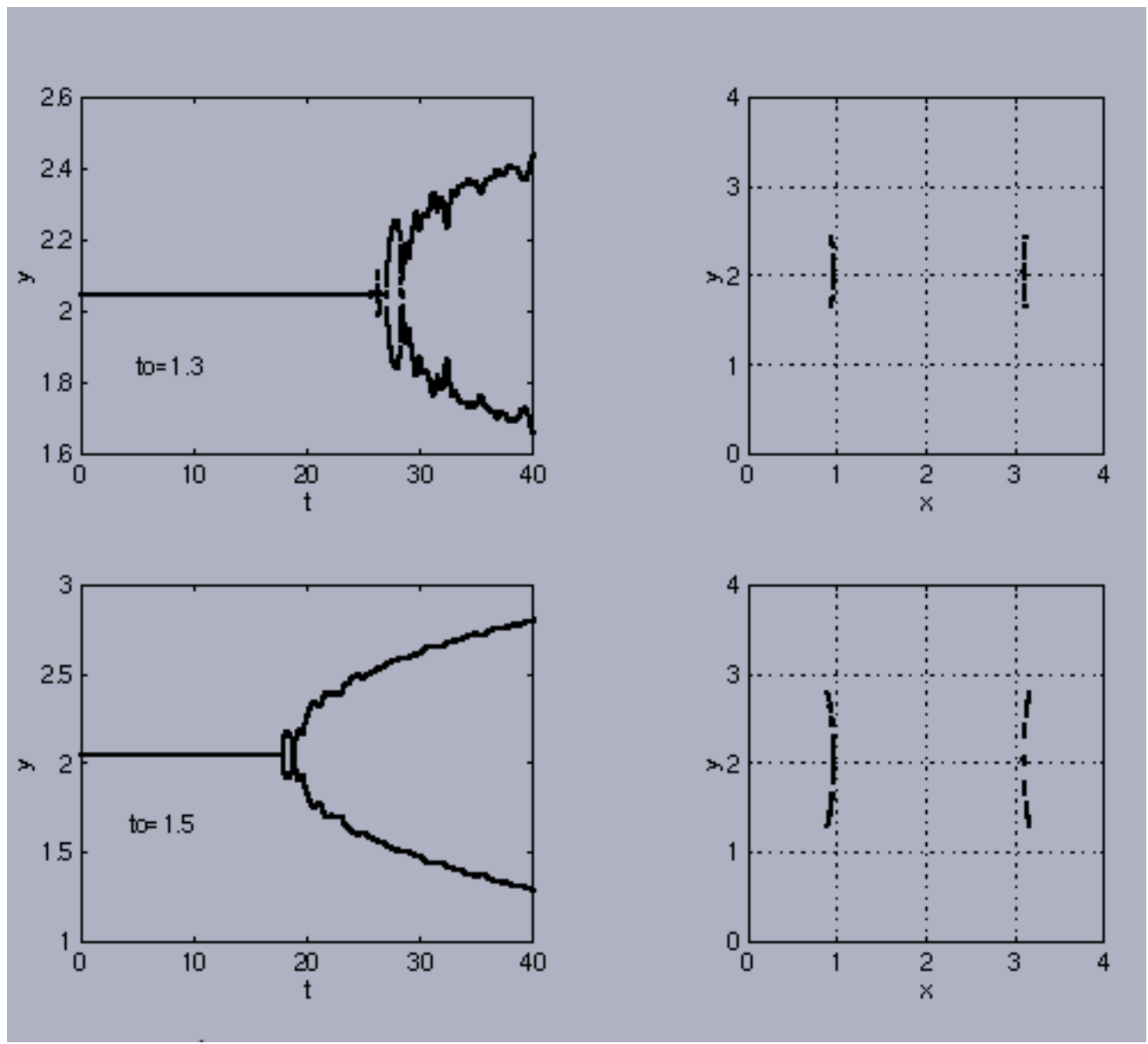}}
\caption{The $CP^1$ solitons $b=(-1,0)$ are run with $\theta_1 \neq 0$ (Skyrme
format)  up to $t=t_0$, at which time $\theta_1$ is set to zero (pure format).
As expected, the lumps divide up more readily for larger $t_0$. The change in
the vertical coordinate of $E_max$ as time goes by, and the trajectory plots
for the two cases $t_0=1.3, 1.5$ are shown.}
\label{fig:pf3}
\end{figure}   

\begin{figure}
\mbox{\epsfig{file=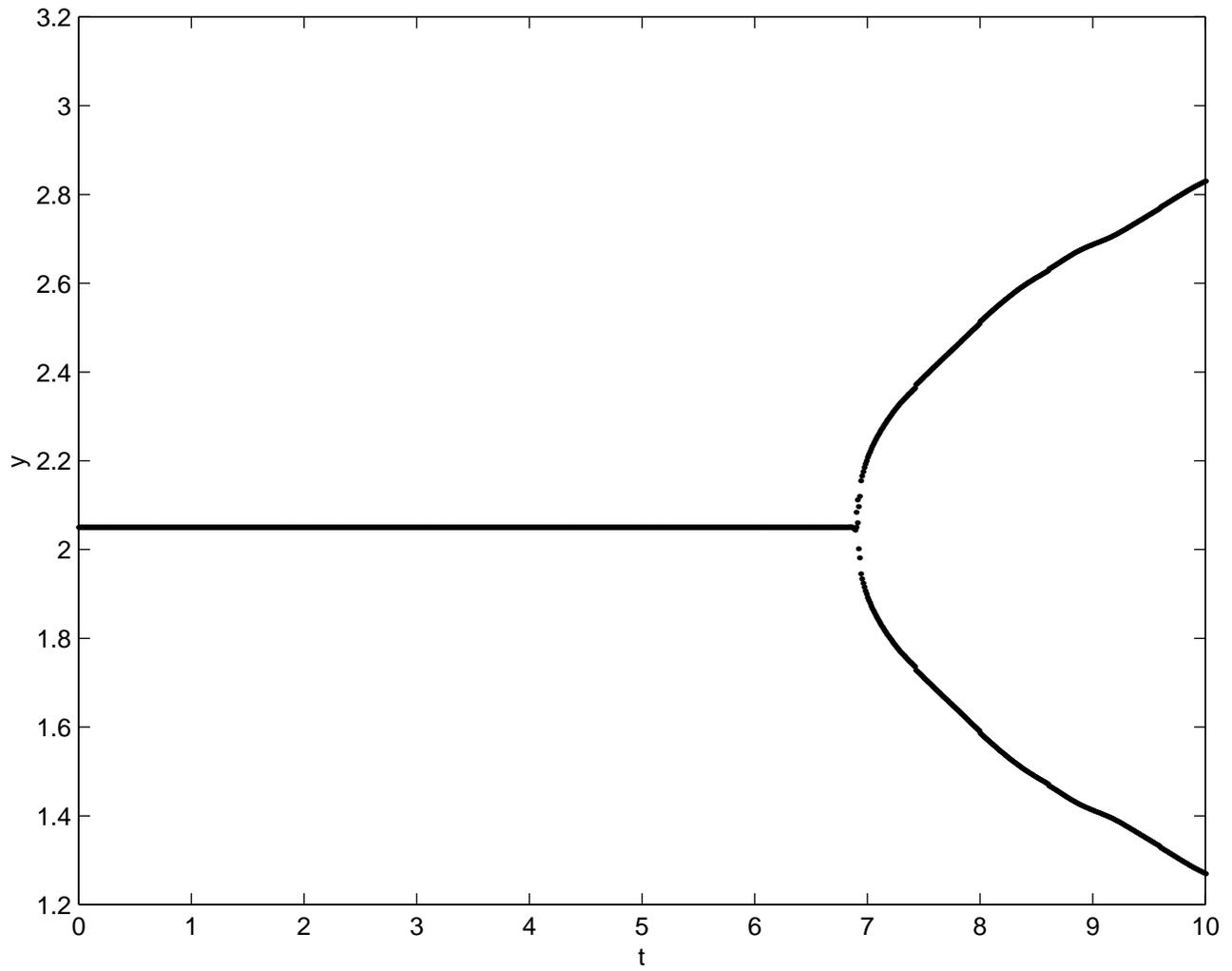}}
\caption{Featuring the division of solitons for $b=(-1,0)$ as in figure 
\ref{fig:pf3}, but with the Skyrme term on at all times.}
\label{fig:pf4}
\end{figure}    

\begin{figure}
\mbox{\epsfig{file=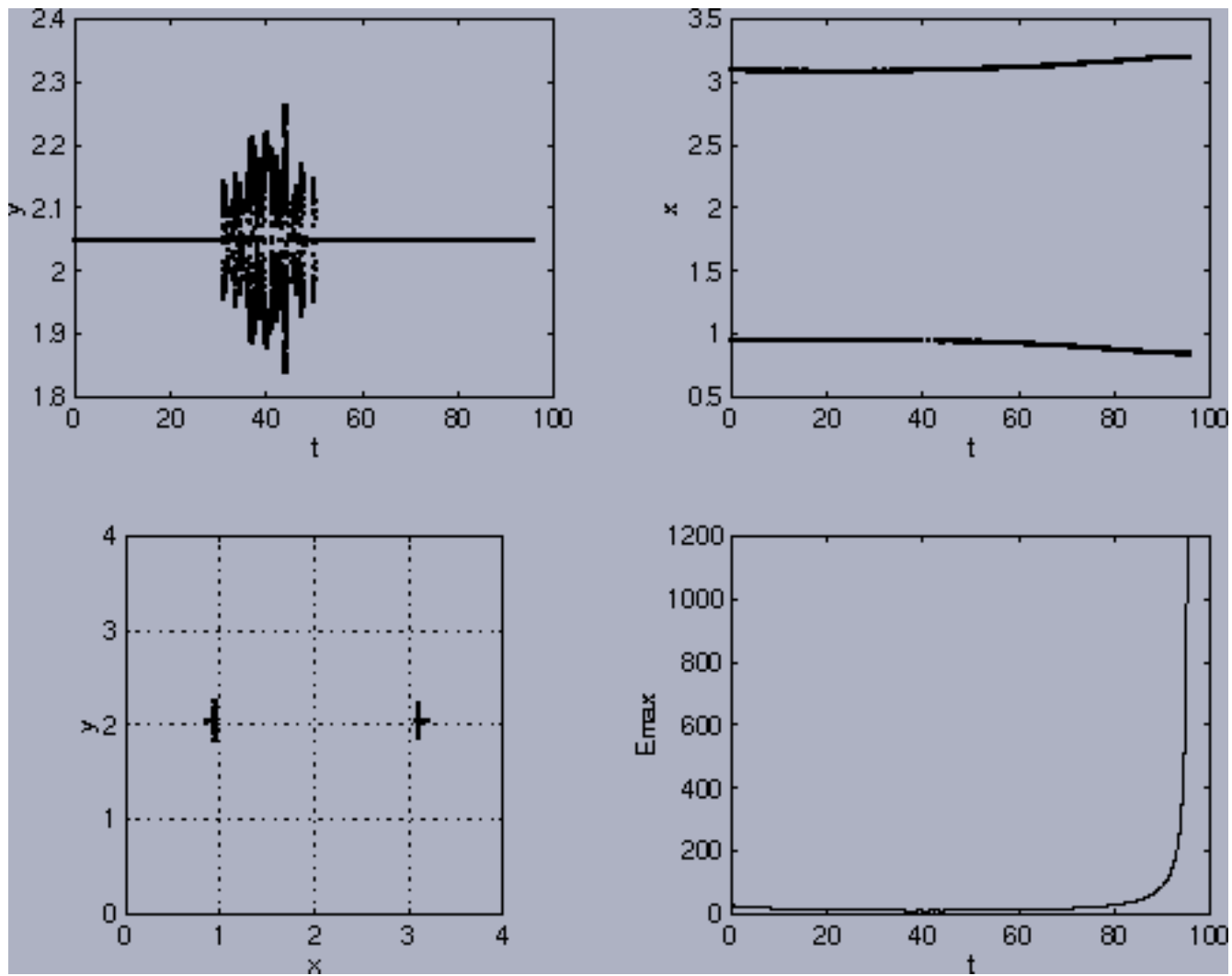}}
\caption{For the critical time $t_0 \approx 1.25$ the energy lumps 
of figure \ref{fig:pf3} can be forced back together.}  
\label{fig:pf5}
\end{figure}        

\begin{figure}
\mbox{\epsfig{file=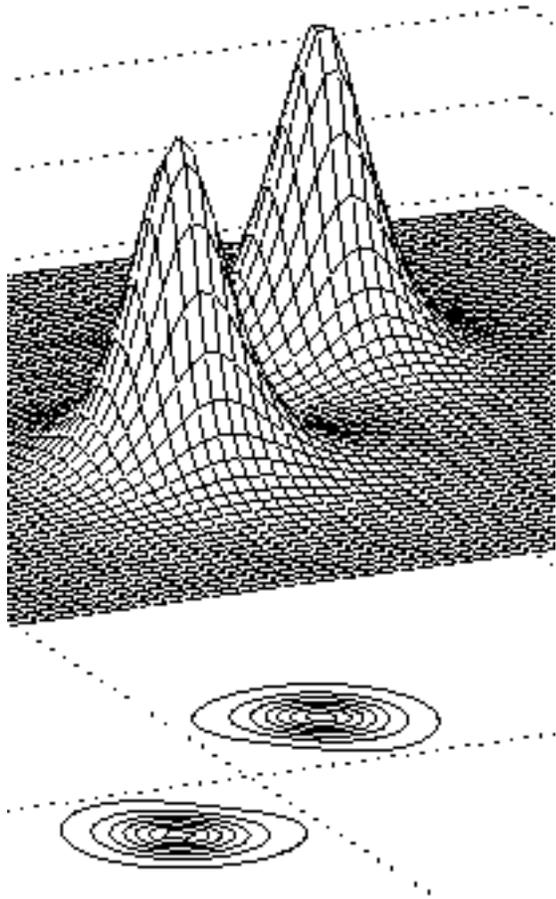}}
\caption{Energy distribution, equation (\ref{eq:energydensity}), for $b$=(0,1).
The extended structures sit along the main diagonals of the lattice. Starting
off from rest, these entities attract each other, collide at the centre of the
grid and scatter at right angles with respect to the initial direction of
motion. (See figure \ref{fig:pf7}.)}
\label{fig:pf6}
\end{figure}

\begin{figure}
\mbox{\epsfig{file=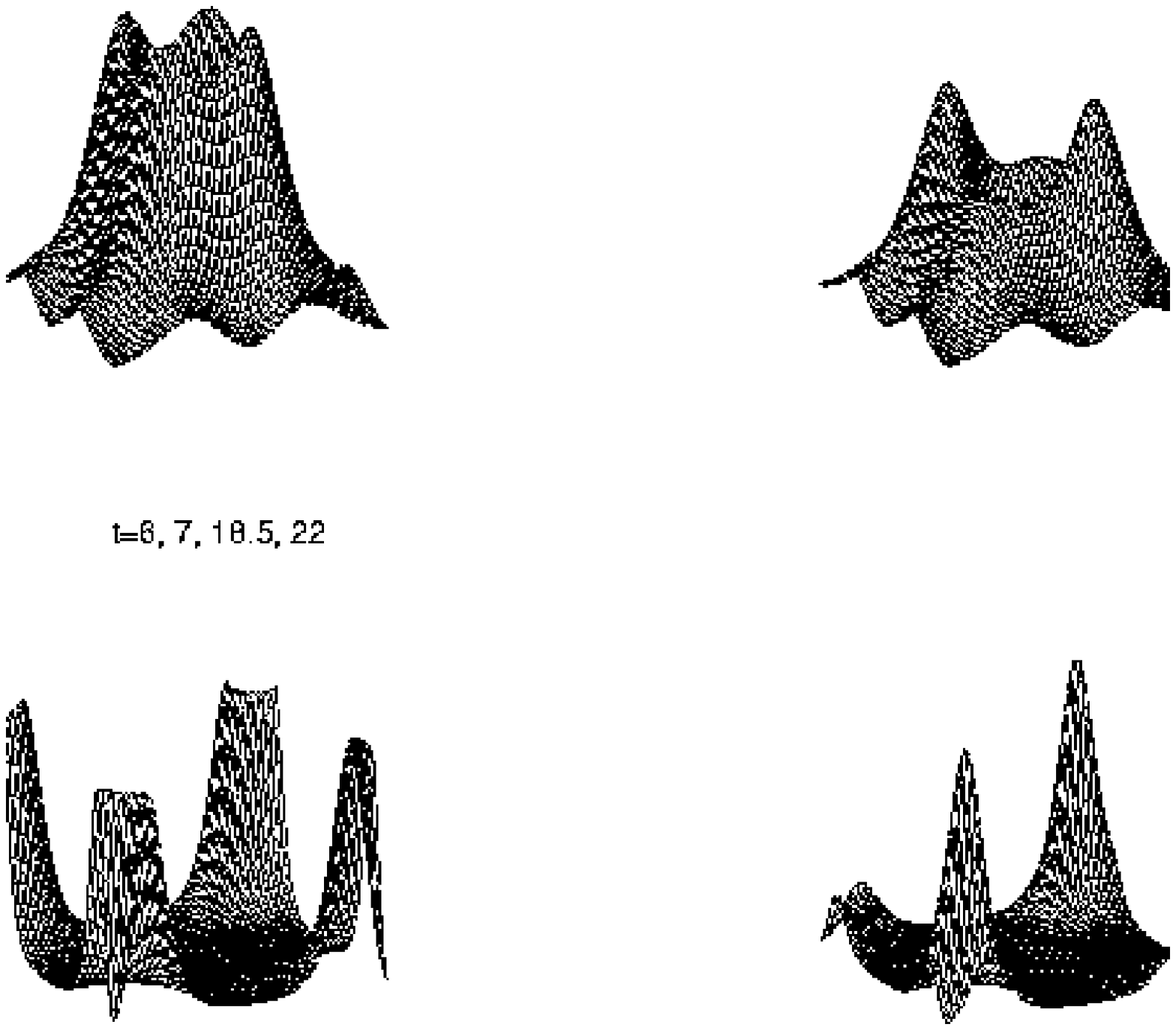}}
\caption{The solitons of picture \ref{fig:pf6} collide at the centre ($t$=6)  
and scatter off at right angles ($t=$7). A similar event occurs at the corners
($t=18.5$), after which the solitons proceed back to the centre. Such
scatterings take place time and again, in periodic cycles. No splitting is
detected.}
\label{fig:pf7}
\end{figure}

\begin{figure}
\mbox{\epsfig{file=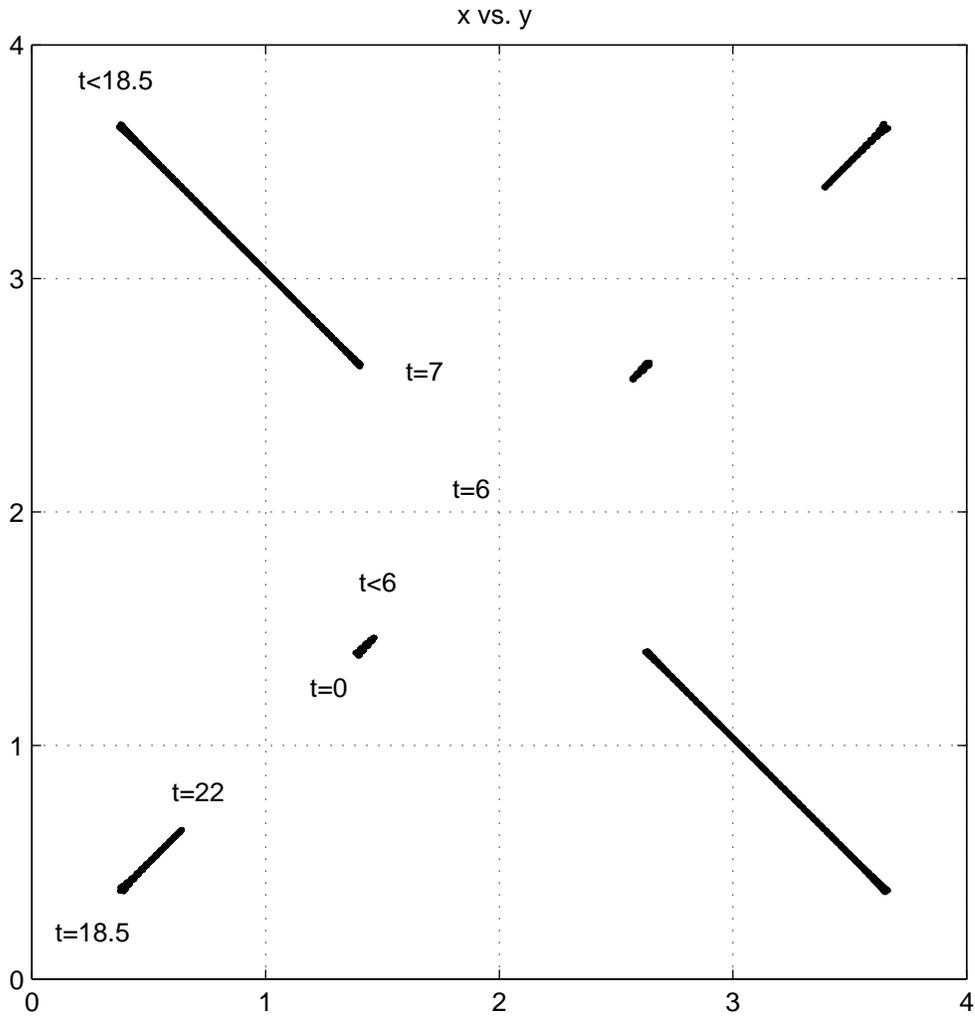}}
\caption{Trajectory plot further illustrating the $b$=(0,1) 
lumps described in the previous picture, figure \ref{fig:pf7}.
The Skyrme term introduces forces 
that make the lumps cruise along the diagonal and collide head-on ($t$=6). The 
scattering at ninety degrees is apparent: Around $t$=18.5 the skyrmions bump into
each other at the corners $(x,y)$=(0,4),(4,0), re-emerging perpendicularly 
through $(x,y)$=(0,0),(4,4).}
\label{fig:pf8}
\end{figure}

\begin{figure}
\mbox{\epsfig{file=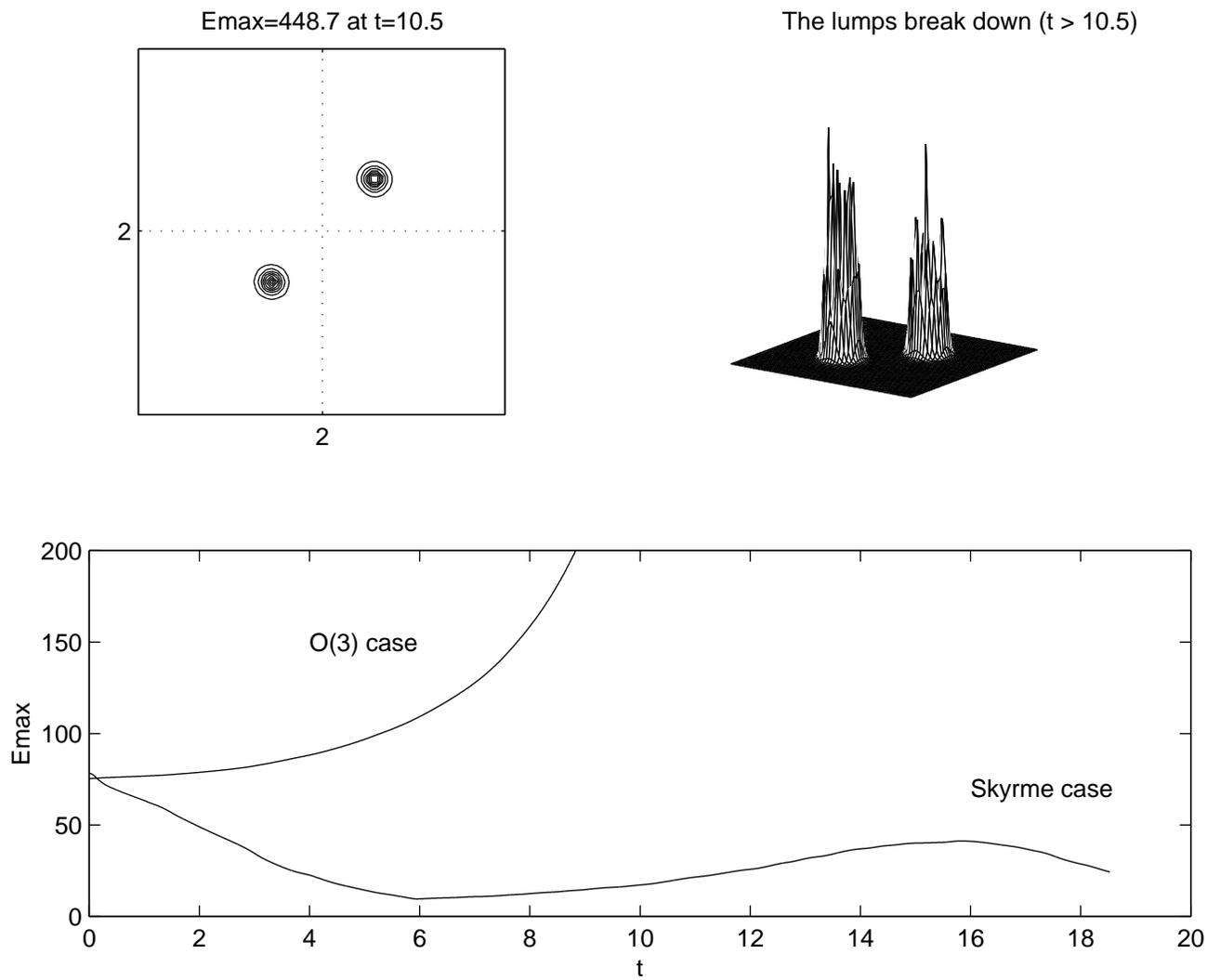}}
\caption{Pure $CP^1$ case corresponding to the value $b$=(0,1). The lumps stay 
still at their initial positions (top-left) as time elapses. They
break down as illustrated at the top-right diagram. Below we have the evolution of
$E_{max}$, including the Skyrme case.}
\label{fig:pf9}
\end{figure}

\begin{figure}
\mbox{\epsfig{file=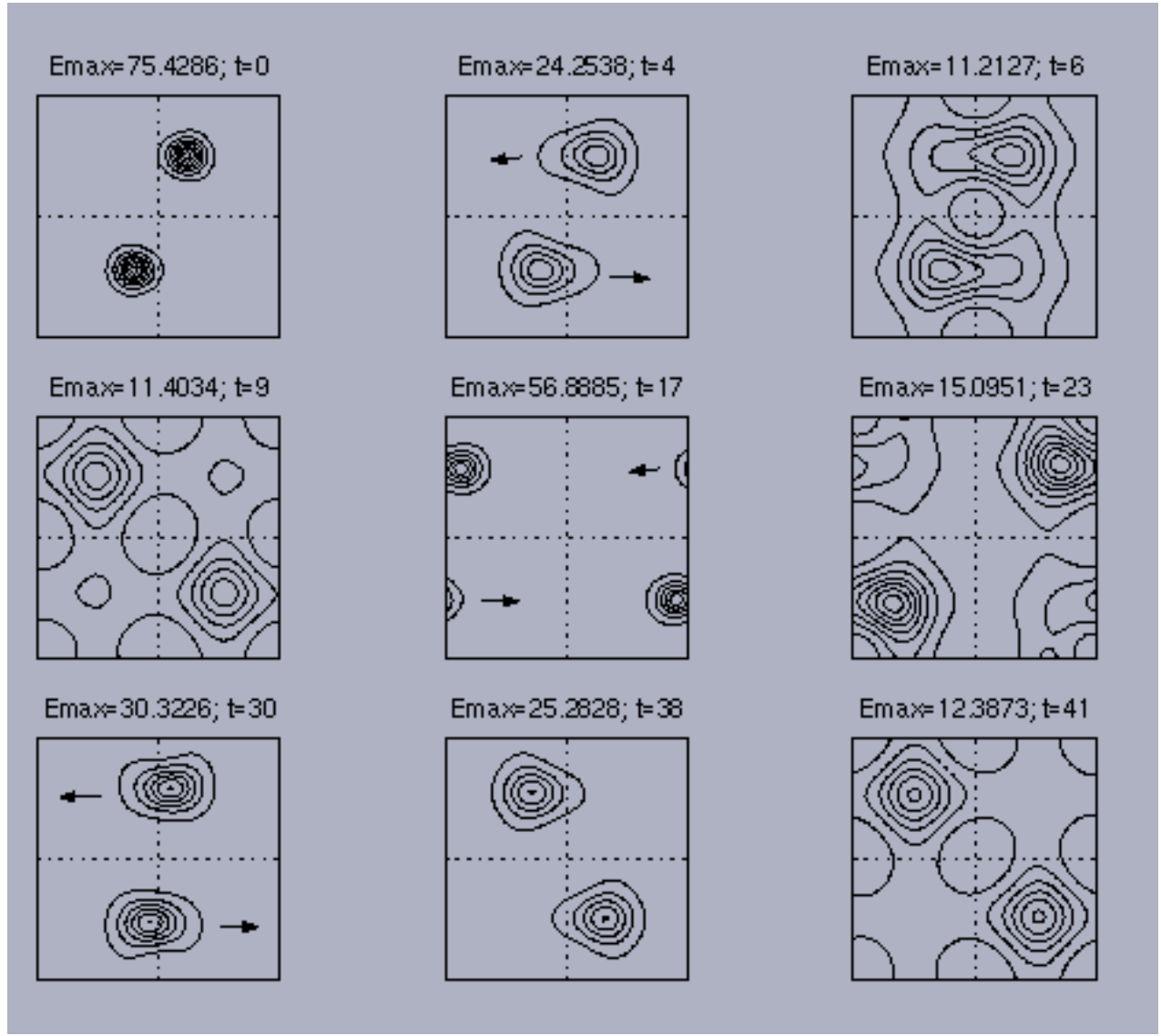}}
\caption{Showing the dynamics arising from $b=(1,1)$. Although no initial speed
is impinged on this system, the skyrmions move resembling a collision at a certain
impact parameter. One skyrmion moves towards decreasing $x$ whereas
the other skyrmion proceeds in the sense of increasing $x$. The 
disappearance/appearance of the lumps through the edges of the lattice reflects the 
periodicity of the network (\ref{eq:cell}).} 
\label{fig:pf10}
\end{figure}

\begin{figure}
\mbox{\epsfig{file=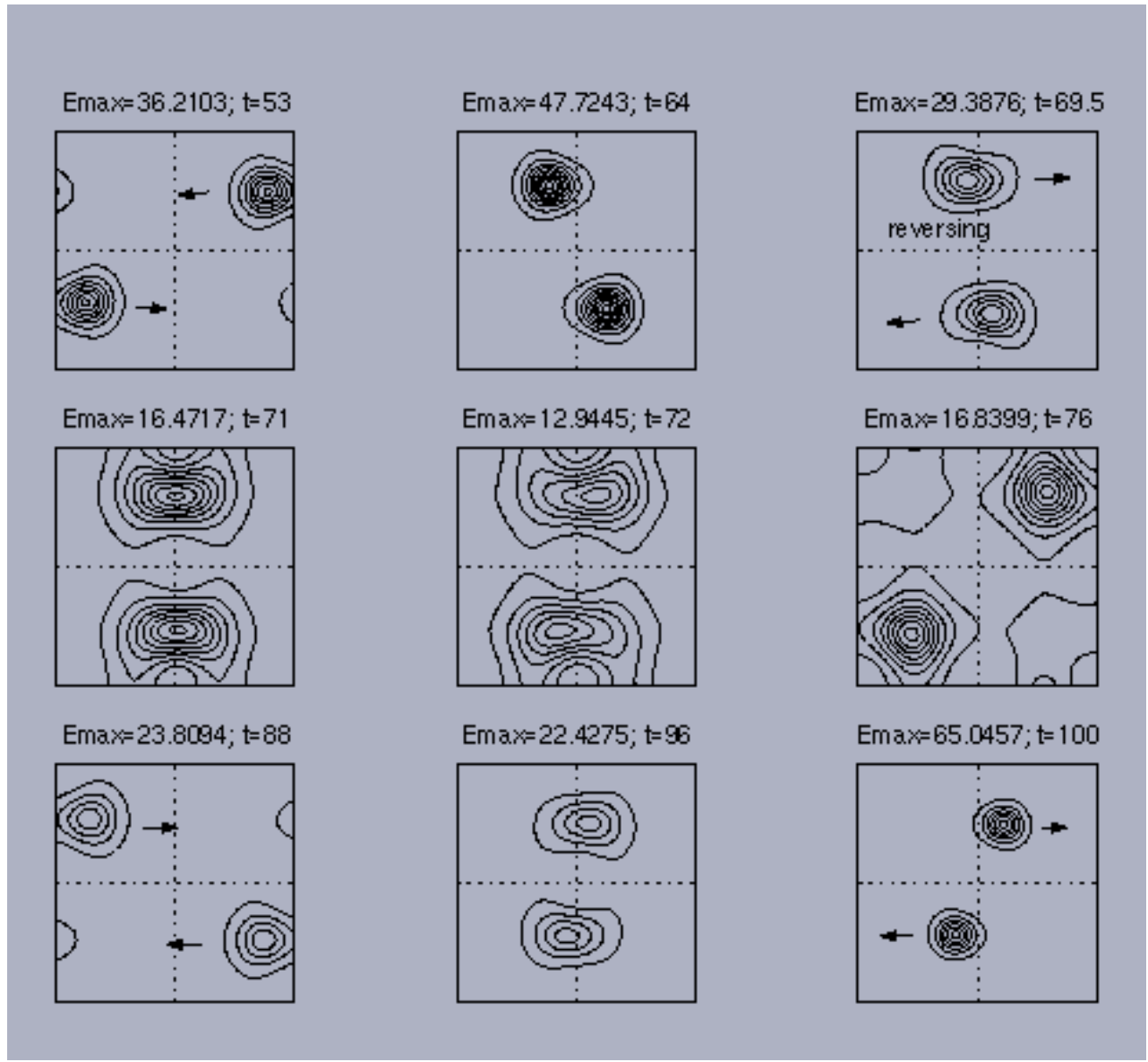}}
\caption{Motion reversal: The skyrmionic structures of figure \ref{fig:pf10} 
continue their itinerary but, at $t=69.5$, their motion begins to reverse.}
\label{fig:pf11}
\end{figure}         

\begin{figure}
\mbox{\epsfig{file=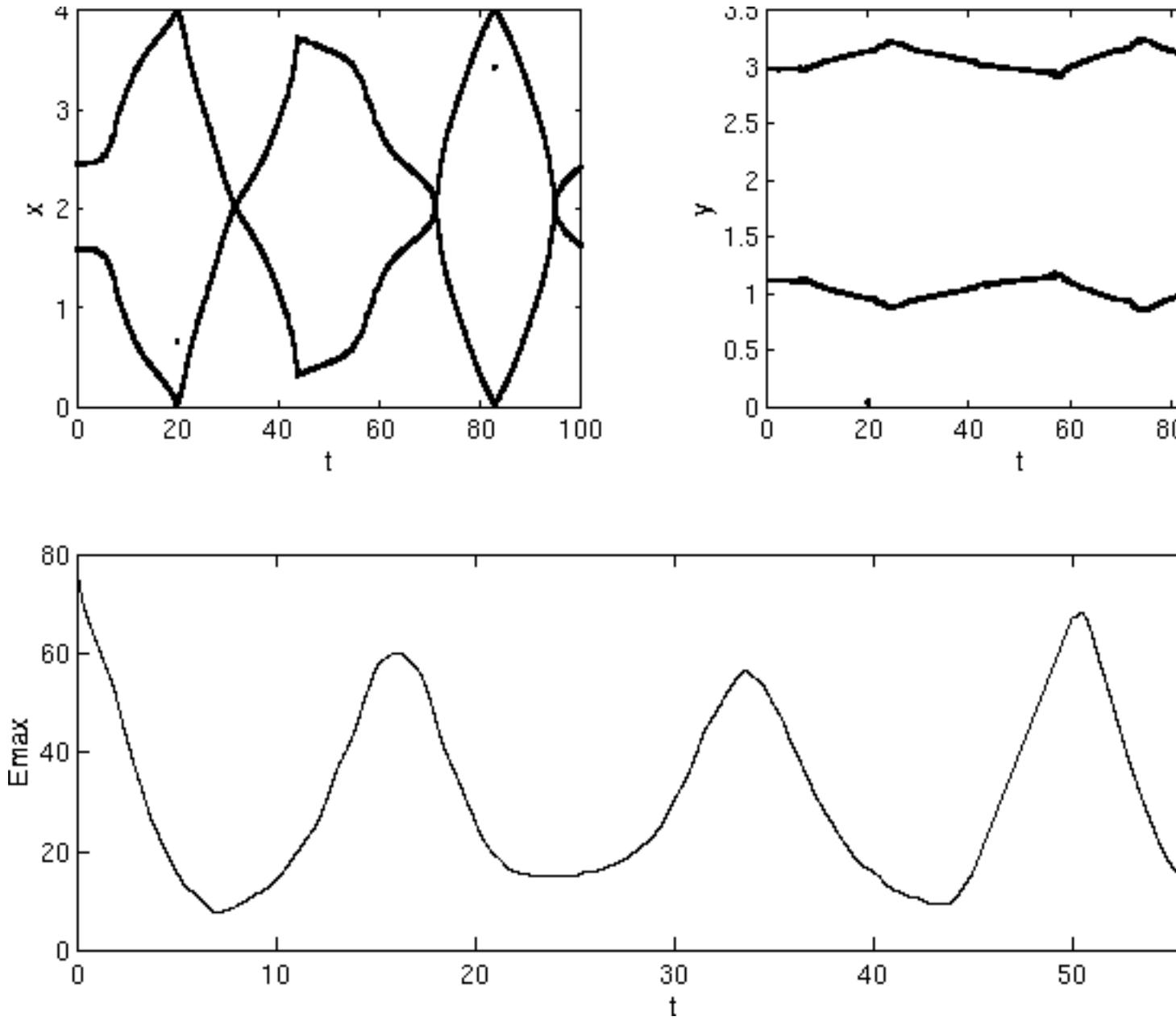}}
\caption{Above: The evolution of the coordinates of $E_{max}$ corresponding to 
the event depicted in figures \ref{fig:pf10} and \ref{fig:pf11}. Below: The
evolution of $E_{max}$ itself.}
\label{fig:pf12}
\end{figure} 

\begin{figure}
\mbox{\epsfig{file=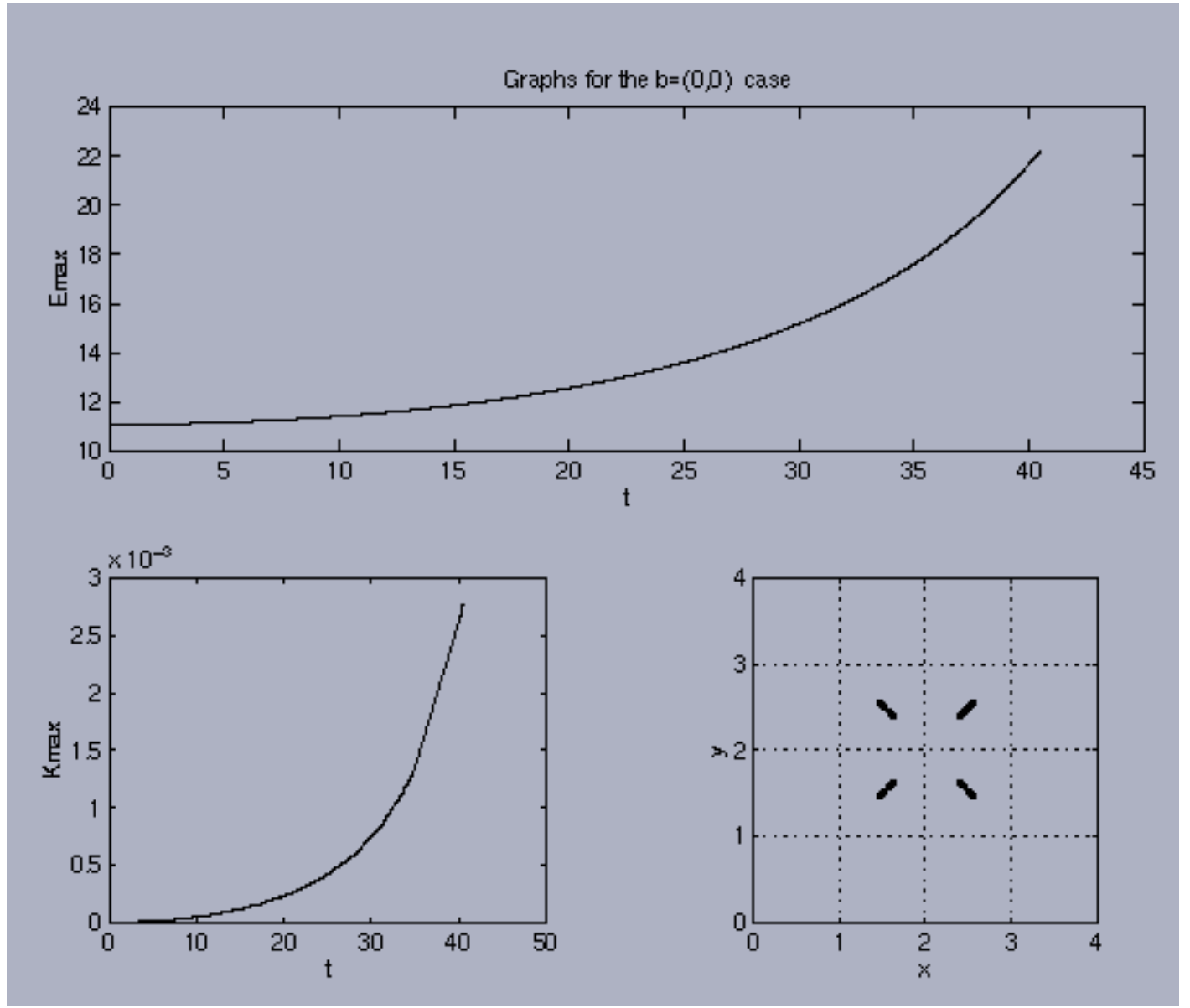}}
\caption{Illustrating the alluring 4-lump, charge-2 state for the
$\theta_1=0$ case. It is an almost stable configuration. The system moves out and 
in along the diagonals oscillating in a breather-like fashion.}
\label{fig:pf13}
\end{figure}

\newpage
\section{Concluding remarks \label{sec:conclusions}}

The $CP^1$ model in (2+1) dimensions has a very rich structure. On the torus we
have uncovered several characteristics qualitatively different from those seen
in the usual model on the compactified plane.  For example in the two skyrmion
sector there exits a configuration featuring four lumps rather than two.
(Clearly, care should now be exercised when referring to the topological charge
as the `lump number'.) We also have the interesting fact that no one soliton
solutions exist on $T_2$. The lumps are not stable in the pure $O(3)$ set-up,
but become stable in the Skyrme model, constructed by the addition of just one
extra term to the $O(3)$ lagrangian.

Skyrmions on $T_2$ have many other important properties. For instance, their
behaviour depends on the elliptic function used to define them.  Thus, skyrmion
fields expressed in terms of the $\sigma$ fields, \emph{e.g.}, equation
(\ref{eq:wsigma}), evolve differently from those expressed through $\wp$,
\emph{e.g.}, equation (\ref{eq:w}). In the former case, energy chunks started
off from rest stay still in their initial positions as time goes by. In the
latter case, we have encountered novel phenomena like lump-splitting,
scattering at ninety degrees and motion reversal, despite no initial speed
being given to the systems. One of the aims of this research has been to
investigate various properties of solitons on $T_2$ by employing alternative
elliptic functions. Comparison of the various solitons generated by alternative
elliptic functions (and with the lumps on $\Re_{2}$) may help us to gain a
deeper insight into the $CP^1$ dynamics.

One of our main results is the existence of a mode of splitting a skyrmion into
two lumps (though we have to work with two skyrmions which then split into four
lumps).  There clearly is a potential barrier which, when overcome, allows the
skyrmions to divide up. We have managed to overcome this barrier by starting
our simulations with a Skyrme initial condition which runs up to a $t=t_0$,
when the Skyrme term is switched off: The evolution continues with
$\theta_1=0$. Such an approach adds a little energy to the system and then
transforms it to the mode responsible for overcoming the barrier. This barrier
is a numerical artifact, brought about by the discretisation procedure.

The longer the system runs with $\theta_1 \ne 0$ the more energy
is transferred; hence there seems to be a minimal time of such
a simulation $t=t_0$ below which
the skyrmions do not break up. This we have seen in our simulations
-- our estimates of $t_0$ gave us $t_0 \sim 1.25$.             

In the present paper we have also learned that splitting takes place only when
the parameter $b$ in $W=\wp+b$ satisfies \( b \in \Re - \{0\} \). For some
values of $b$ we have found that the forces operating in $\wp$-lump systems
with $v_0=0$ either lead to head-on collisions and a subsequent 90$^{\circ}$
scattering or lead to a motion reversal. This property is qualitatively
different to any other $CP^1$ structures that we know of.

\bigskip
The table below summarises some of our results:

\bigskip
%
\begin{center}
\begin{tabular}{|l|c|c|r|} \hline
$Im(b)/Re(b)$       & = 0                      & $\neq 0$  \\ \hline
=0                  & 4 lumps on diagonals;    & 2 lumps on central cross; \\
                    & breather                 & splitting  \\ \hline
$\neq$ 0            & 2 lumps on diagonals;    & 2 lumps elsewhere  \\
                    & 90$^{\circ}$ scattering    & motion reversal    \\  \hline
\end{tabular}
\end{center}            

\bigskip
In future work we hope to report on:
\begin{itemize}
\item configurations with $v_0 \neq 0$ on the central cross to further analise 
the splitting phenomenon;
\item lumps initially situated slightly
off the central cross \( [b=(\alpha,\beta), \; 0 < \beta \ll 1] \);
\item lumps slightly off the main diagonals
      \( [b=(\alpha,\beta), \; 0 < \alpha \ll 1] \);
\item obtain deeper insight into motion reversal.
\end{itemize}

\bigskip
A look a bit farther ahead will necessarily include configurations with
higher Brouwer degree, as well as  states defined through other
functions, \emph{e.g.}, the elliptic ones of Jacobi and the
pseudo-elliptic theta functions.  

\newpage 
\Large{\bf Acknowledgements} \\
\normalsize

Part of this work was carried out when $\Re$J Cova was on a research visit at
the Department Mathematical Sciences of the University of Durham, and later
when WJZ visited the Physics Department of La Universidad del Zulia. We thank
the Royal Society of London and Venezuela's Conicit for supporting these
visits. We are also grateful to both Universities for their hospitalities.

\smallskip
CONDES project 01797-00 is also thankfully acknowledged.



\begin{thebibliography}{99}
%
\bibitem{cho} Proceedings \emph{`Physics in (2+1) dimensions' (Korea),
World Scientific} (1992)
%
\bibitem{skyrme} Skyrme THR \emph{Nucl. Phys.} {\bf 31}, 556 (1962)
%
\bibitem{witten83} Witten E \emph{Nucl. Phys. B} {\bf 223}, 422 (1983)
%
\bibitem{chaos} Piette B and Zakrzewski W \emph{Chaos, solitons and fractals} 
{\bf 5}, 2495 (1995)
%
\bibitem{non} Cova RJ and Zakrzewski WJ {\em Nonlinearity} {\bf 10}, 1305 (1997)
%
\bibitem{split} Cova RJ \emph{Helv Phys Acta} \textbf{71}, 675 (1998)
%
\bibitem{martin} Speight JM \emph{Comm Math Phys} \textbf{194}, 513 (1998)
%
\bibitem{goursat} Goursat E \emph{Functions of a complex variable},
Dover Publications (1916)
%
\bibitem{lawden} Lawden DF {\em Elliptic functions and applications}, 
Springer Verlag (1989)
%
\bibitem{private} Martin JM, \emph{private communication}
%
\bibitem{ejp} Cova RJ \emph{Euro. Phys. J.} \textbf{B 15}, 
673 (2000)
%
\bibitem{rosen} Rosenzweig C and Mohan Srivastava A \emph{Phys. Rev.} 
\textbf{D 43} 12, 4029 (1991)
%
\bibitem{atiyah} Atiyah M and Hitchin N \emph{The geometry and dynamics of 
magnetic monopoles}, Princeton University Press (1988)
%
%
\bibitem{samols} Samols T \emph{Comm Math Phys} \textbf{145}, 149 (1992)
%
\end{thebibliography}
\end{document}